\documentclass[12pt,preprint]{aastex}
\usepackage{graphicx}

\slugcomment{Accepted for the ApJ Letters, June 24th 2004}

\shorttitle{SMA sub-mm continuum observations of Orion-KL}
\shortauthors{Beuther et al.}

\begin{document}


\title{Sub-arcsecond sub-mm continuum observations of Orion-KL}


\author{H. Beuther\altaffilmark{1},
Q. Zhang\altaffilmark{1}, 
L.J. Greenhill\altaffilmark{1,2}, 
M.G. Reid\altaffilmark{1}, 
D. Wilner\altaffilmark{1},
E. Keto\altaffilmark{1},
D. Marrone\altaffilmark{1}, 
P.T.P. Ho\altaffilmark{1}, 
J.M. Moran\altaffilmark{1},
R. Rao\altaffilmark{1}, 
H. Shinnaga\altaffilmark{1}, 
S.-Y. Liu\altaffilmark{3}}

\altaffiltext{1}{Harvard-Smithsonian Center for Astrophysics, 60 Garden Street, Cambridge, MA 02138, USA}
\altaffiltext{2}{Kavli Institute of Particle Astrophysics and Cosmology,
SLAC, 2575 Sand Hill Rd, Menlo Park, CA 94025.}
\altaffiltext{3}{Academia Sinica Institute of Astronomy and Astrophysics,  No.1, Roosevelt Rd, Sec. 4, Taipei 106, Taiwan, R.O.C.}

\email{hbeuther@cfa.harvard.edu}







\begin{abstract} We present the first $865\,\mu$m
continuum image with sub-arcsecond resolution obtained with the
Submillimeter Array. These data resolve the Orion-KL region into the
hot core, the nearby radio source I, the sub-mm counterpart to the
infrared source n (radio source L), and new sub-mm continuum
sources. The radio to submillimeter emission from source I may be
modeled as either the result of proton-electron free-free emission
that is optically thick to $\sim 100$\,GHz plus dust emission that
accounts for the majority of the submillimeter flux, or H$^-$
free-free emission that gives rise to a power-law spectrum with
power-law index of $\sim 1.6$. The latter model would indicate similar
physical conditions as found in the inner circumstellar environment of
Mira variable stars. Future sub-arcsecond observations at shorter
sub-mm wavelengths should easily discriminate between these two
possibilities. The sub-mm continuum emission toward source n can be
interpreted in the framework of emission from an accretion disk.
\end{abstract}





\section{Introduction}

In spite of being the nearest (450\,pc) and most studied region of
massive star formation, we do not understand Orion-KL adequately. The
best known source in Orion-KL is the BN object, a heavily reddened B0
star that may be a run-away star from the Trapezium cluster
\citep{plambeck1995,tan2004}. The region exhibits a complex cluster of
infrared sources studied from near- to mid-infrared wavelengths
\citep{dougados1993,greenhill2004,shuping2004}. At least two outflows
are driven from the region on scales $>10^4$\,AU, one high-velocity
outflow in the south-east north-west direction observed in molecular
lines and in the optical and near-infrared (e.g.,
\citealt{allen1993,wright1995,chernin1996,schultz1999}), and one lower
velocity outflow in the north-east south-west direction best depicted
in the thermal SiO and H$_2$O maser emission as well as some H$_2$ bow
shocks (e.g.,
\citealt{genzel1989,blake1996,chrysostomou1997,stolovy1998}). The
driving source(s) of the outflows are uncertain: initial claims that
it might be IRc2 are outdated now, and possible culprits are the radio
sources I and/or the infrared source n, also known as radio source L
\citep{menten1995}.

Radio source I lies close to the center of a biconical outflow on the
order of $10^3$\,AU across, that is traced by SiO and H$_2$O maser
emission \citep{gezari1992,menten1995,greenhill2003} and has not been
detected in the near- to mid-infrared \citep{greenhill2004}. Its
spectral energy distribution from 8 to 86\,GHz can
be explained by optically thick free-free emission. The
turnover frequency has not been observed yet
\citep{plambeck1995}. However, the fact that SiO masers, typically
associated with luminous (evolved) stars such as Mira variables, have
been detected toward source I indicates that different physical
processes might take place (e.g., \citealt{menten1995}). To
get a better idea about the emission process and physical nature of
source I, continuum observations at higher frequency are needed.  So far
this has been an ambitious task because the peak of the hot core is
only $1''$ east of source I and dominates the region unless observed
with sub-arcsecond resolution \citep{plambeck1995}.

\section{Observations and data reduction}
\label{obs}

Orion-KL was observed with the Submillimeter Array (SMA\footnote{The
Submillimeter Array is a joint project between the Smithsonian
Astrophysical Observatory and the Academia Sinica Institute of
Astronomy and Astrophysics, and is funded by the Smithsonian
Institution and the Academia Sinica.}) on February 2nd 2004 at
348\,GHz ($865\,\mu$m) with 7 antennas in its so far most extended
configuration with baselines between 15 and 205\,m. The phase center
was the nominal position of source I as given by \citet{plambeck1995}:
R.A. [J2000] $5^{\circ}35'14''.50$ and Dec. [J2000]
$-5^{\circ}22'30''.45$. For bandpass calibration we used the planets
Jupiter and Mars. The flux scale was derived by observations of
Callisto and is estimated to be accurate within 15\%. Phase and
amplitude calibration was done via frequent observations of the quasar
0420-014 about 17$^{\circ}$ from the phase center. The zenith opacity
measured with the NRAO tipping radiometer located at the Caltech
Submillimeter Observatory was excellent with $\tau(\rm{348GHz})\sim
0.125$. The receiver operates in a double-sideband mode with an IF
band of 4-6\,GHz. The correlator has a bandwidth of 2\,GHz and the
spectral resolution was 0.825\,MHz corresponding to a velocity
resolution of 0.7\,km\,s$^{-1}$. System temperatures were between 250
and 600\,K. The $1\sigma$\,rms in the final image is 35\,mJy, mainly
determined by the side-lobes of the strongest source, the hot core.
The synthesized beam is $0.78''\times 0.65''$. We calibrated the data
within the IDL superset MIR developed for the Owens Valley Radio
Observatory and adapted for the SMA, the imaging was performed in
MIRIAD. For more details on the array and its capabilities see the
accompanying paper by \citet{ho2004}.

On the shortest baselines, there is nearly no line-free part in the
2\,GHz spectral window (see also \citealt{schilke1997b}), whereas on
the longest baselines only the strongest lines remain. Taking spectra
toward selected sources (I, n, hot core, SMA1), we find that only the
outflow tracers~-- thermal SiO and SO$_2$~-- are strong toward the
sources I and n. Emissions from other species are concentrated toward
the hot core and SMA1 and are not observed toward the sources I and n
(the line data will be published elsewhere).  Therefore, to construct
a pseudo-continuum from the spectral line data we excluded the
strongest lines from the spectrum (e.g., SiO, SO$_2$), and averaged
the rest of the spectrum (about 1.65\,GHz) into a continuum
channel. We estimate that the derived sub-mm continuum fluxes of the
sources I and n are accurate within the calibration uncertainty of
15\%, whereas the other sources in the field are contaminated by line
emission, and thus their measured fluxes are upper limits.

\section{Results}

\begin{figure}
\includegraphics[angle=-90,width=8cm]{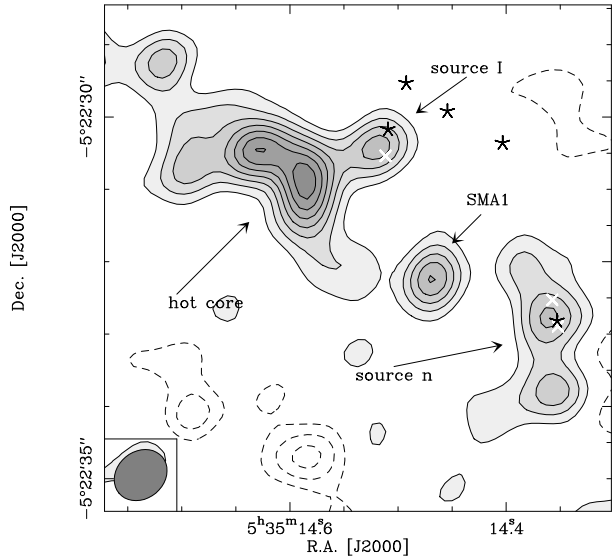}
\caption{Sub-mm continuum image of the Orion-KL region at
865\,$\mu$m. The contouring starts at the $2\sigma$ level of
70\,mJy/beam and continues in $2\sigma$ steps. Positive emission is
shown in grey-scale with contours, negative features~-- mainly due to
missing flux~-- are presented in dashed contours. The white crosses
mark the radio positions of sources I and n (two crosses for n because
of the bipolar structure) \citep{menten1995}.  The black stars mark
the positions of emission peaks at 3.8\,$\mu$m \citep{dougados1993},
rotated to the J2000 frame and corrected to account for the proper
motion of BN (e.g., \citealt{plambeck1995,tan2004}), which was used as
an astrometry reference in near-infrared observations (position
uncertainties are $0.03''-0.05''$). The beam is shown at the bottom
left ($0.78''\times 0.65''$).}
\label{continuum}
\end{figure}

The integrated flux within the field mapped by the SMA is only a few
Jy, whereas the flux measured with single-dish instruments is $\sim
170$\,Jy \citep{schilke1997b}. Therefore, we filter out more than 90\%
of the flux of the region and just sample the most compact components
within the massive star-forming cluster (Figure
\ref{continuum}). Clearly, we distinguish source I from the hot
core. The positional offset between the sub-mm peak and the radio
position of source I from \citet{menten1995} is $\sim 0.1''$ which is
about our calibration uncertainty. Source I is the only source still
detected on the longest baselines and thus must be extremely compact.
Even the intrinsically strongest source, the hot core, vanishes at the
largest baselines ($>100$\,k$\lambda$). Furthermore, we detect a
sub-mm counterpart to the infrared and radio source n, and another new
source approximately between the sources I and n which we label SMA1
(the fluxes of all sources are given in Table \ref{fluxes}). The image
of SMA1 as well as source n is sensitive to the chosen
uv-range. However, using all data of this observation with different
weighting and data reduction schemes both features are consistently
reproducible. Comparing our data with mm continuum images of the
region \citep{plambeck1995,blake1996}, we do detect source n in the
sub-mm band although it was not detected previously at mm wavelengths.
The morphology of the hot core is similar in the previous mm
observations and the new sub-mm continuum data. However, both mm
datasets with lower spatial resolution only show a small elongation
toward SMA1 whereas we resolve it as a separate emission peak at
865\,$\mu$m with sub-arcsecond resolution. It is possible that SMA1 is
embedded within the larger scale hot core which is filtered out by the
extended array configuration we use. Therefore, it is difficult to
judge whether SMA1 is a separate source of maybe protostellar nature
or whether it is just another peak of the hot core ridge.

\section{Discussion}

\begin{figure}
\includegraphics[angle=-90,width=8cm]{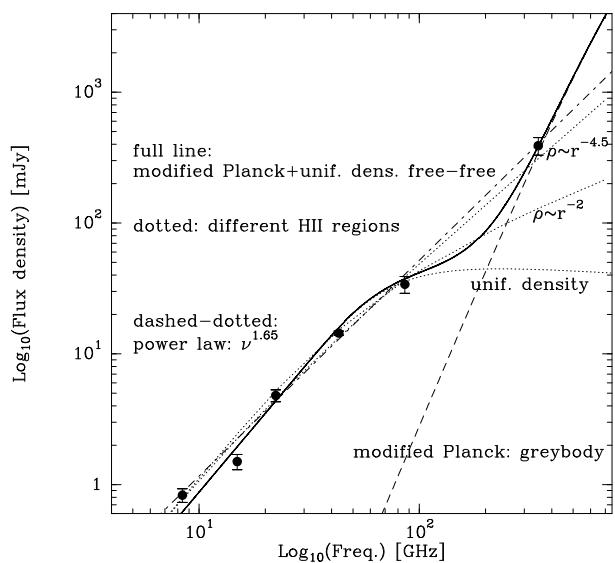}
\caption{Spectral energy distribution of source I: the measured fluxes
are labeled as dots with error-bars, and the various lines show
different possibilities to fit the data as labeled within the
plot. The 15, 22 and 86\,GHz fluxes are taken from
\citet{plambeck1995} and references therein, the 8 and 43\,GHz fluxes
are more recent values (consistent within the error-bars of the previous
measurements by \citealt{menten1995}, Reid et al. in prep.).}
\label{sed}
\end{figure}

\underline{\it Source I:} The brightness temperature of the continuum
emission from source I at 43\,GHz ranges from 1600 K at the emission
peak to about 800\,K near the south-east and north-west edges of the
source (Menten et al. in prep.).  This could either be optically thin
emission from gas at $\sim 10^4$\,K, where hydrogen is ionized
(proton-electron free-free), or partially optically thick emission
from gas at $\sim 1600$\,K, where atomic and molecular hydrogen is
neutral and electrons come from the partial ionization of metals
(H$^-$ and H$_2^-$ free-free; these terms are used even though the
interactions of the hydrogen and electrons do not involve bound states
of negative ions).  The latter case applies to the radio photospheres
of Mira variables at roughly 2 stellar radii
\citep{reid1997}. Figure \ref{sed} shows the spectral energy
distribution (SED) for source I from 8 to 348\,GHz. Two
interpretations of the data are possible.

{\it Proton-electron free-free+dust emission:} As we are dealing
with a young massive star-forming region, one can fit the
lower-frequency part of the spectrum with proton-electron free-free
emission (from now on labeled as free-free emission), whereas in the
sub-mm band protostellar dust emission starts to dominate the spectrum
(e.g., \citealt{hunter2000}). Figure \ref{sed} shows three different
SEDs for the free-free emission with various density distributions
that can fit the data from 8 to 86\,GHz. The models with density
gradients were done within the procedures outlined in
\citet{keto2002,keto2003}. While the model with uniform density yields
approximately constant flux at frequencies greater than 100\,GHz, the
free-free fluxes further increase in the sub-mm band for H{\sc ii}
regions with density gradients. The uniform density model allows us to
calculate the ultra-violet photon flux from the optically thin
emission and thus to estimate the luminosity of the source to $\sim
10^{3.6}$\,L$_{\odot}$ (see, e.g, \citealt{spitzer1998}). This is
consistent with the dynamical mass estimate $\leq 10$\,M$_{\odot}$
\citep{greenhill2003}, corresponding to $L \leq
10^{3.76}$\,L$_{\odot}$. In addition, based on the SiO maser emission,
\citet{menten1995} state that the source is likely to be rather
luminous (probably $\geq 10^{4}$\,L$_{\odot}$). With the data so far,
it is difficult to discriminate between the uniform density model and
models with density gradients. However, there appears to be excess
flux at 348\,GHz that is likely due to optically thin dust
emission. We can estimate the dust contribution by using the uniform
density H{\sc ii} region model for a lower limit of the free-free
contribution $S_{\rm{free-free}}\geq 44$\,mJy, which results in an
upper limit for the dust contribution of $S_{\rm{dust}}\leq
276$\,mJy. Assuming a dust temperature of 100\,K and a dust opacity
index $\beta$ of 2 (a lower $\beta$ results in too much dust
contribution at lower frequencies and degrades the fits), we estimate
the resulting gas mass and gas column density from source I to
$M\rm{_{\rm{gas}}} \leq 0.2\,\rm{M}_{\odot}$ and $N\rm{_{\rm{gas}}}
\leq 8.5\times 10^{24}\,\rm{cm}^{-2} $ (for more details on the
assumptions and range of errors see, e.g.,
\citealt{hildebrand1983,beuther2002a}.).  This upper limit to the gas
mass within the potential circumstellar disk is about an order of
magnitude below the approximate dynamical mass of source I of the
order 10\,M$_{\odot}$, \citep{greenhill2003}. This is different from
disk studies at the earliest evolutionary stages of massive star
formation where estimated disk masses are of the same order as the
masses of the evolving massive stars (e.g., IRAS\,20216+4104,
\citealt{zhang1998a}). From an evolutionary point of view, this
implies that source I should be more evolved than IRAS\,20126+4104. In
spite of the low gas mass, we find high column densities corresponding
to a visual extinction $A_{\rm{V}}$ of the order 1000. The extinction
toward source I is significantly higher than the $A_{\rm{V}}\sim 60$
toward the close by source IRc2 derived from the infrared data
\citep{gezari1992}. However, very high extinction is necessary to
explain the the non-detection of source I in the near- and
mid-infrared as well as the X-ray band
\citep{dougados1993,greenhill2004,garmire2000}. The column densities
have to be at distances from the center between 25\,AU (outside the
SiO maser emission) and 320\,AU (the spatial resolution of the
observations).

{\it H$^-$ and H$_2^-$ free-free:} One can also fit a power law
$S\propto \nu ^{\alpha}$ to the SED with $\alpha \sim 1.65\pm
0.2$. This is similar to the spectral index observed toward Mira
variable stars \citep{reid1997}. Evidence that the radio continuum
forms under Mira-like conditions in a region with a temperature $\sim
1600$\,K and a density of $10^{11-12}$\,cm$^{-3}$ comes from the
detection of SiO masers from source I.  The v=1 J=1-0 (43\,GHz) SiO
maser emission is from the first vibrationally excited state at $\sim
1800$\,K above the ground-state, and models of maser pumping require
temperatures of roughly 1200\,K and hydrogen densities of the order
$10^{9-10}$\,cm$^{-3}$ for strong maser action \citep{elitzur1992}.
Since the continuum emission requires only $\sim 400$\,K higher
temperature and perhaps a factor of 10 higher density than the SiO
masers, the near juxtaposition of these two emitting regions could be
just as in Mira atmospheres. The path length needed to achieve H$^-$/H$_2^-$
free-free optical depth unity for material at a density of
$10^{11}$\,cm$^{-3}$ and temperature $\sim 1600$\,K is $\sim 2$\,AU.
The observed spectral index at cm wavelengths is slightly under 2 for
both Miras and source I, suggesting roughly similar variations of
opacity with radius.  This model has the benefit that the a single
power-law can explain the observed spectral energy distribution
between 8 and 350\,GHz.

The disk observed toward source I extends roughly $0.05''$ (25\,AU)
from the star \citep{greenhill2003}. If we assume that there is strong
opacity at infrared wavelengths (i.e., optically thick near the peak
of a 1600\,K blackbody) then the luminosity of the source will be
given by $L\propto \rm{area_{surface}} \times \sigma \times T^4$. For
a thin disk with a temperature of 800\,K at the outer radius of 25\,AU
and an assumed typical temperature profile of $T\propto r^{-0.5}$
between 4 and 25\,AU from the center (e.g., \citealt{reid1997}), the
luminosity is $\sim 2 \times 10^4$\,L$_{\odot}$.  This is consistent
with source I being a luminous object (of the order
$10^4$\,L$_{\odot}$, \citealt{menten1995}) but not exceeding the total
luminosity of the KL region ($L\sim 10^5$\,L$_{\odot}$,
\citealt{genzel1989}). The dynamical mass of source I is estimated to
$\leq 10$\,M$_{\odot}$ \citep{greenhill2003}, corresponding to $L \leq
10^{3.8}$\,L$_{\odot}$. Thus a simple, consistent case can be made
that the continuum emission from source I has H$^-$/H$_2^-$ free-free
as the dominant source of opacity and that there is a transition
between a disk photosphere and the SiO maser emission at a radius of
$\sim 25$\,AU.

{\it Solving the problem:} While observations at 230\,GHz might
already help in discriminating between both scenarios, the flux
differences are more obvious at higher frequencies
(Fig. \ref{sed}). The free-free + dust emission models predict
690\,GHz fluxes $\sim 3.7$\,Jy, whereas the H$^-$/H$_2^-$ free-free
model predicts fluxes on the order 1.2\,Jy. SMA observations at
690\,GHz should easily discriminate between both scenarios.

\underline{\it Source n:} We detect a sub-mm counterpart to source n
plus an adjacent sub-mm source about $1''$ to the
south. Morphologically, these two sub-mm sources resemble the bipolar
radio structure observed by \citet{menten1995}, but the radio
structure is on smaller scales ($0.4''$) within the sub-mm counterpart
of source n. The southern sub-mm source and the northern elongation of
the dust emission of source n follow approximately the direction of
the H$_2$O maser outflow and the bipolar radio source. It is tempting
to associate these features as potentially caused by the outflow, but
as we cannot set tighter constraints we refrain from this
interpretation. Source n is detected at a 1\,mJy level at 8\,GHz and
not detected down to a threshold of 2\,mJy at 43\,GHz
\citep{menten1995}. Assuming that the cm flux is due to free-free
emission, its contribution at 348\,GHz is negligable. Therefore, the
observed sub-mm flux likely stems from optically thin dust
emission. Employing the same assumptions as for source I, we again can
calculate the gas mass and column density to $M\rm{_{\rm{gas}}} \sim
0.27\,\rm{M}_{\odot}$ and $N\rm{_{\rm{gas}}} \sim 5.7\times
10^{24}\,\rm{cm}^{-2}$.  There is still no general consensus as to
which sources are the driving engines of the outflows observed in
H$_2$O emission, and both sources, I and n, are possible
candidates. Recently, extended mid-infrared emission was observed
toward source n perpendicular to the outflow axis
\citep{greenhill2004,shuping2004} which is interpreted as possibly due
to an accretion disk. Furthermore, \citet{luhman2000} detected CO
overtone emission toward source n and interpreted that also in the
framework of an irradiated disk. In this scenario, the sub-mm
continuum emission stems from this disk, and thus the derived gas mass
of $\sim 0.27$\,M$_{\odot}$ could correspond to the mass of this
potential accretion disk. The mid-infrared observations of source n
indicate a non-edge-on orientation of the disk
\citep{shuping2004,greenhill2004}. The new sub-mm continuum data, by
themselves are consistent with a high column density through the disk
are consistent with the inclination scenario.

\underline{\it The hot core and SMA1:} The hot core splits up into at
least three sources over a region of about 1000\,AU. In comparison,
\citet{greenhill2004} report that the infrared source IRc2 also splits
up into a similar amount of sources on small spatial scales indicating
possible very high source densities.  However, it is not settled yet
whether the hot core and/or IRc2 are internally or externally
heated. In the latter case, the hot core is just the remnant of the
dense core from which the other sources have formed.  The source SMA1
is associated with strong CH$_3$CN emission (to be presented
elsewhere) and has no cm or infrared counterpart. Assuming that this
condensation is protostellar in nature it would be one of the youngest
sources in the whole Orion-KL region. In addition, the location of
SMA1 between source I and n is intriguing, and one has to take this
source into account as a possible driving source of one or the other
outflow in the KL region. However, as it is close to the hot core and
shows similar emission line signatures, one can also associate it with
the hot core and thus question whether the source is externally or
internally heated. As the derived flux values of the hot core and SMA1
are line contaminated we refrain from a further interpretation of
these parameters.

\acknowledgments{We like to acknowledge the tremendous work of the
whole SMA staff for making this instrument possible! Thanks a lot also
to the referee Dr. Antonio Chrysostomou for valuable comments
improving the quality of the paper.  H.B. acknowledges financial
support by the Emmy-Noether-Program of the Deutsche
Forschungsgemeinschaft (DFG, grant BE2578/1).}


\clearpage

\begin{deluxetable}{lrrrrr}
\tablecaption{865\,$\mu$m parameter and results\label{fluxes}}
\tablewidth{0pt}
\tablehead{
\colhead{Source} & \colhead{$S_{\rm{peak}}$} & \colhead{$S_{\rm{int}}^{c}$} & \colhead{$T_{\rm{b}}$} & \colhead{$M_{\rm{gas}}$} &  \colhead{$N_{\rm{gas}}$}\\
\colhead{} & \colhead{mJy\,beam$^{-1}$} & \colhead{mJy} & \colhead{K} & \colhead{M$_{\odot}$} & \colhead{cm$^{-2}$}
}
\startdata
Source I  & 320 & 320$^a$ & 6.2 & 0.2 & 8.5\,10$^{24}$\\
Source n  & 300 & 360 & 7.0 & 0.27 &  5.7\,10$^{24}$\\
SMA1$^b$ & 360 & 360$^a$ & 7.0 & -- & -- \\
Hot core$^b$  & 540 & 1870 & 10.5 & -- & -- \\
\enddata
\tablenotetext{a}{\footnotesize Unresolved}
\tablenotetext{b}{\footnotesize Probably line contamination (\S \ref{obs})}
\tablenotetext{c}{\footnotesize Integrated emission}
\end{deluxetable}

\clearpage

\end{document}